For a German language PDF version of the article, click here.

# THE REDSHIFT OF EXTRAGALACTIC NEBULAE

## (Die Rotverschiebung von extragalaktischen Nebeln)

## Fritz Zwicky

(Translated by H. Andernach (heavily editing a raw translation from https://tradukka.com/translate)

**Abstract:** This gives a description of the most essential characteristics of extragalactic nebulae, as well as of the methods used to investigate these. In particular, the so-called redshift of extragalactic nebulae is discussed in detail. Various theories which have been proposed to explain this important phenomenon, are briefly discussed. Finally, it will be indicated to what extent the redshift promises to become of importance for the study of cosmic rays.

**Table of Contents**

---

Special terms were interpreted as follows:

- Nebel = nebula (not galaxy, to preserve the original flavor of the text)

- Fluchtgeschwindigkeit = recession velocity (not escape velocity, as there is no escape here...)

- Perioden-Helligkeits-Beziehung = Period-Brightness Relationship (not period-luminosity, to preserve the original word "Helligkeit")

- penetrating radiation = cosmic rays

# 1. INTRODUCTION

It has been known for a long time that there are certain objects in space, which appear, when observed with small telescopes, as very blurry, self-luminous patches. These objects possess structures of different types. Often they are spherical in shape, often elliptical, and many of them have a spiral-like appearance, which is why they are occasionally called spiral nebulae. Thanks to the enormous angular resolution of modern giant telescopes, it was possible to determine that these nebulae lie outside the bounds of our own Milky Way. Images taken with the 100-inch telescope on Mt. Wilson reveal that these nebulae are stellar systems similar to that of our own Milky Way. By and large, the extragalactic nebulae are distributed uniformly over the sky and, as has been demonstrated, are also distributed uniformly in space. They occur as single individuals or group themselves to clusters. The following lines intend a brief summary of the more important characteristics and a description of the methods that made it possible to establish these characteristics.

# 2. DISTANCES AND GENERAL CHARACTERISTICS OF EXTRAGALACTIC NEBULAE

As already mentioned, it is possible, with the help of modern telescopes, to resolve a number of nebulae. wholly or partially, into single stars. In the great Nebula in Andromeda, for example, a great number of individual stars have been observed. Recently also globular star clusters have been discovered in this nebula, similar to those which lie within our own Milky Way. The fortuitious fact of the observability of individual stars in nebulae opens two ways to determine their distances.

**A) Distance Determination with the help of the Period-Brightness Relationship for Cepheids.**

Cepheids are stars the brightness of which varies periodically with time. Periods are usually in the range of one up to sixty days. The absolute magnitude is a unique function of period, a function which has been determined for the stars of our own system [Milky Way]. If the period is known, it is therefore possible to derive the absolute magnitude of these cepheids from this relationship. If, in addition to that, one determines the apparent magnitude, and compares it with the absolute magnitude, one immediately obtains the distance of the stars. Quite a few cepheids have been observed in the Andromeda nebula. Based on these, the distance and diameter of Andromeda was determined to be 900,000 and 42,000 light-years, respectively. For comparison, it is important to remember that our own system has a diameter the upper limit of which is estimated to be about 100,000 light years. The distances of eight other nebulae have been found in the same way. In nebulae more distant than a few million light years, individual cepheids cannot be resolved. To determine their distance other methods must therefore be devised.

**B) Statistics of the brightest stars of a nebula**

This method is based on the assumption that in the extragalactic stellar systems the relative frequency of the absolute magnitude of stars is the same as in our own system. The experience with the previously examined neighboring systems is in fact in accordance with this assumption. The absolute magnitude of the brightest stars in our own and neighboring systems turns out to be -6.1 on average, with a dispersion of less than half a magnitude. We just note that similar distance determinations were obtained with the help of novae.

**C) Distance determination of nebulae using their total apparent magnitude.**

With the help of the first two methods, the distances of about sixty extragalactic nebulae have been found. From the measured apparent brightness and the known distance of these nebulae we can immediately infer their absolute brightness. In this way we obtain the following distribution curve (Fig. 1).

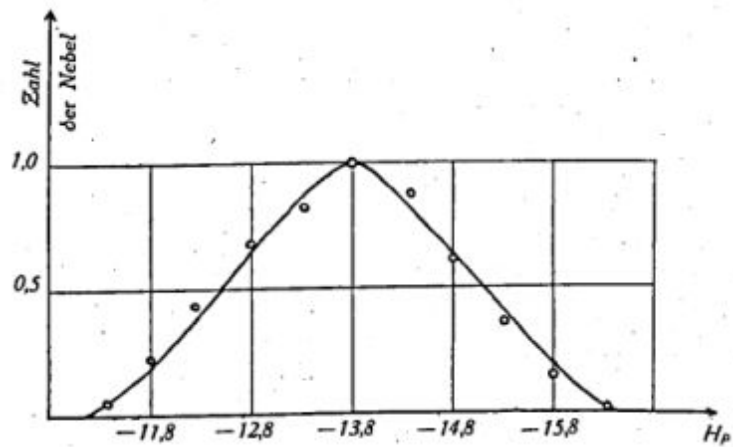

**Figure 1.** $H_P$ = absolute photographic magnitude (Y axis: number of nebulae).

The average absolute visual magnitude of the nebulae is -14.9 with a dispersion of about five magnitudes and a half-width of the distribution curve of about two magnitudes. This dispersion is unfortunately too big to allow an exact determination of the distance to an individual nebula from its apparent brightness and the distribution curve of absolute magnitudes. We shall discuss later how it is still possible to determine the distance to certain individual nebulae with great accuracy. However, the following fact allows us to find the distance of a large number of extremely faint nebulae. As already mentioned, nebulae often group themselves in dense clusters, which contain from 100 to 1000 individuals. It is of course extremely likely that such an apparent accumulation of nebulae is also a real accumulation in space, and that therefore all these nebulae are located at approximately the same distance. It is relatively easy to determine the distribution curve of apparent magnitudes of the nebulae of a cluster. This distribution curve is virtually the same as the distribution curve of the absolute magnitudes of the sixty nebulae, whose distances have been found under (A) and (B). This proves that the apparent accumulation of nebulae [on the sky] corresponds to a real dense swarm in outer space. A comparison of mean apparent brightness of the nebulae in the cluster with the mean absolute magnitude of -14.9 immediately yields the distance of the cluster. The distances of the following clusters of nebulae were determined in this way.

**Table 1.** Distance in millions of light years of various galaxy clusters

| | |
|---|---:|
| Coma-Virgo | 6 |
| Pegasus | 23.6 |
| Pisces | 22.8 |
| Cancer | 29.3 |
| Perseus | 36 |
| Coma | 45 |
| Ursa Major I | 72 |
| Leo | 104 |
| Gemini | 135 |

The number of nebulae per unit volume in one of these dense swarms is at least a hundred times greater than the corresponding average number of individual nebulae dispersed in space.

It is of interest to include here some brief comments regarding other features of nebulae which are accessible to research with the help of the 100-inch telescope.

With regard to the structure of the Universe, first and foremost is the question whether the distribution of nebulae over space is uniform or not. In the case of uniformity we expect the number of nebulae in a spherical shell of radius $r$ and constant thickness $dr$ to be proportional to $r^2$, provided that we are dealing with a Euclidean space. This expectation actually corresponds very accurately to reality, i.e. for that part of the Universe within reach of the 100-inch telescope. This does not mean, of course, that space may not eventually turn out as non-euclidian, once we are able to penetrate farther in space.

We must not fail to mention that the above conclusions are only valid in case that absorption and scattering of light in space may be ignored. The finding of a uniform distribution of nebulae to the largest achievable distances with a method that assumes the practical lack of absorption and scattering, is in fact by itself almost a proof for the correctness of this assumption. Indeed, an actually existing uniform distribution of nebulae would be biased by absorption, in such a way that the number of nebulae in spherical shells of constant thickness would increase more weakly with distance than $r^2$, and eventually even decrease. In view of the fact that gases and clumps of dust can be proven to exist in interstellar space of our system [Milky Way], it would nevertheless be of great importance to have an independent proof of the transparency of intergalactic space, and to show that it is not the curvature of space, combined with absorption and scattering, that would feign a uniform distribution of the nebulae. A statistical study of the apparent diameters of nebulae would, for example, serve this purpose.

Theoretically, the presence of intergalactic matter should correspond to the vapor pressure of the extant star systems. Assuming that the Universe has reached a steady state, it is possible to estimate this pressure (F. Zwicky, Proc. of the Nat. Academy of Sci., vol. 14, p. 592, 1928). It turns out to be extremely small and would practically exclude the detection of interagalactic [The German original says "intragalactic", but the translator supposed that this was a typographical error] matter.

Another interesting question is related to the spectral types of nebulae. Most extragalactic nebulae possess absorption spectra similar to that of the Sun with strong salient H and K lines of calcium and an intense G-band of Ti (4308 Å), Fe (4308 Å) and Ca (4308 Å). Therefore, nebulae belong to the spectral type G. The spectral type is independent of the distance, as far as current observations reach out. A distance-dependent displacement of the total spectrum will be discussed later. The width of the absorption lines is usually several Angstroms and is also independent of the distance.

A small percentage of the observed nebulae also show emission lines (Nebulium), usually originating in the core region of the nebulae. Unfortunately very little is known as yet about the physical conditions in such systems.

Thirdly, it is of importance to investigate the relative frequency of the already mentioned different forms of nebulae. The statistical distribution is approximately 74% spirals, 23% spherical nebulae, and about 3% show an irregular appearance.

Fourthly, I would like to mention the determination of the brightness distribution within a single nebula. This investigation has been undertaken recently by E. Hubble at Mt. Wilson. Hubble obtains the following preliminary result. The brightness can be expressed as a universal function $L(r, \alpha)$, where $r$ is the distance from the center of the nebula, and $\alpha$ is an appropriate parameter. By varying $\alpha$, one can reduce the brightness distributions in all nebulae with great accuracy (approximately 1%) to the same function, in fact up to values of $r$, for which the brightness has fallen to 1/1000 of that of the centre. It is also of importance with respect to the practical lack of absorption and scattering in intergalactic space, that the distribution function of the $\alpha$'s of the different nebulae is independent of distance. Incidentally, we mention that $L$ coincides with the function that corresponds to the brightness distribution in an isothermal Emden gas sphere.

Fifth, it is of enormous importance that the nebulae at a great distance show redshifted spectra, where the shift increases with distance. The discussion of the so-called redshift is the main topic of the present work.

# 3. THE REDSHIFT OF EXTRAGALACTIC NEBULA. RELATIONSHIP BETWEEN DISTANCE AND REDSHIFT

V. M. Slipher at the Observatory in Flagstaff, Arizona, was the first to observe that some nebulae show shifts of their spectra, which correspond to a Doppler effect of up to 1800 km/s. However, Slipher did not establish any relationship between redshift and distance. Such a relationship was first suspected by G. Strömberg (1925, ApJ 61, 353-388) upon his study of the speed of the Sun relative to more and more distant objects. He found that the mean velocity of the Sun, relative to the system of neighboring nebulae, is large, of the order of 500 km/s, and that the group of the nebulae used shows an expansion which seems to depend on the distance of the individual nebula.

Since at the time of Strömberg's research no reliable distance determination of the nebula was known, K. Lundmark attempted to relate the observed high velocities with the compactness of the photographic images of the nebula. This proved later as an attempt in the right direction. Nevertheless, the attempt remained unsuccessful, as it turned out that the apparent diameter of nebulae at the same distance exhibit large variations.

E. Hubble worked at Mt. Wilson in the same direction. First he also tried to relate the redshift to the apparent concentration of the nebula. Herein he set out from the idea that the redshift would correspond to the well-known Einstein effect. However, it turned out that it was not possible to uncover sensible relationships in this way.

Consequently, E. Hubble tried to relate the redshift with the distance of the different nebulae. This attempt, as is well known, has since been of great success. The nebulae which were available initially for such an investigation, had distances from one up to six million light years. The discussion of all data showed a linear relationship between redshift and the distance, with the result that the redshift correspondeds to an apparent recession velocity of 500 km/s per one million parsec (1 parsec equal to approximately 3.26 light years). The dispersion was however relatively large, as for example the neighboring Andromeda Galaxy has a violet shift of approx. 200 km/s, i.e. is either seemingly or really moving towards us. Despite this, it was found later that this spectral shift, calculated for the first time, was an extremely good one. The best proof of the amazing care of Hubble's method of work is perhaps, that on the basis of the above relationship up to now he could predict the redshifts in each case to within a few percent, and in fact for distances up to thirty times higher that those of the initially used [sample of nebulae].

The difficulty in photographing the spectra of very distant nebulae lies in the need of extremely long exposure times. Indeed it was necessary to expose plates up to fifty hours and more, and it seemed hardly possible to penetrate further into space with this method. In more recent times great progress has been made by using a spectrograph, the camera lens of which has a focal ratio of f/0.6. However, with this one had to sacrifice a lot in the dispersion, and the spectra obtained are only about 2 millimeters long. However, the exposure times could be shortedened to a few hours. Nevertheless, it does not seem possible to penetrate farther into space than about 200 million light years. The reason for this lies partly in the location of the 100-inch telescope in the vicinity of the large city of Los Angeles, since the illumination of the night sky and the associated strong light scattering into the telescope unfortunately limits the astronomical observations on Mt. Wilson, to a level below the actual performance of the telescope. For the 200-inch telescope, currently under construction for the California Institute of Technology, a more suitable location will therefore have to be selected.

The redshifts of various clusters of nebulae, expressed as apparent Doppler recession velocities, are put together in the following Table 2.

**Table 2.**

| Cluster of nebulae | Number of nebulae in the cluster | Apparent diameter, in degrees | Distance in $10^6$ light years | Average velocity, in km/s |
|---|---|---|---|---|
| Virgo | (500) | 12 | 6 | 890 |
| Pegasus | 100 | 1 | 23.6 | 3810 |
| Pisces | 20 | 0.5 | 22.8 | 4630 |
| Cancer | 150 | 1.5 | 29.3 | 4820 |
| Perseus | 500 | 2.5 | 36 | 5230 |

| Coma | 800 | 1.7 | 45 | 7500 |
| Ursa Major I | 300 | 0.7 | 72 | 11800 |
| Leo | 400 | 0.6 | 104 | 19600 |
| Gemini | (300) | | 135 | 23500 |

These results are graphed in Fig. 2.

(see E. Hubble and M. L. Humason, ApJ 74, 43, 1931. In this work, also the most essential bibliography may be found.)

It follows from this compilation that extragalactic nebulae have velocities which are proportional to their distance. The specific velocity, per million parsec, is

$$v_s = 558 \text{ km/s.} \tag{1}$$

The redshift of each individual nebula is on average based on the mean shifts of at least three spectral lines. These are usually the H- and K-lines, the G-band ($\lambda = 4303$ Å), and occasionally one of the lines Hδ (4101 Å), Hγ (4340 Å), Fe (4384 Å) and Hβ (4861 Å). The uncertainty in the redshift of the cluster of nebulae in Leo results in this way for example, as

$$v = 19621 \pm 300 \text{ km/s.}$$

The various absorption lines suffer the same relative shift, just as expected for the Doppler effect. Thus we have for a certain nebula:

$$\Delta\lambda/\lambda = \text{constant} = K = v/c = \kappa r \tag{2}$$

regardless of the wavelength, $\lambda$, and conveniently the shift may be expressed, as we have done, in velocity units. The same value of K therefore also applies to the shift of the maximum of the continuous emission spectrum.

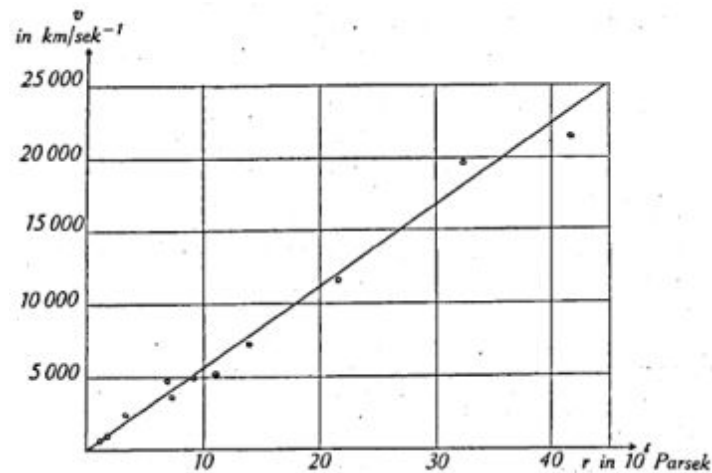

**Figure 2.**

It must be borne in mind that Fig. 1 lists the mean Doppler velocity of clusters of nebulae. This speed is the average of the values of several individual nebulae (from 2 to 9) in each cluster. It is of great importance for the theory of the effects discussed here, that the velocities of the individual members of a cluster can deviate from the mean. In the Coma cluster, for example, which until now is the best investigated, the following individual values have been measured.

Apparent velocities in the Coma cluster

|  |  |
|---|---|
| v = 8500 km/s | 6000 km/s |
| 7900 | 6700 |
| 7600 | 6600 |
| 7000 | 5100(?) |

It is possible that the last value of 5100 km/s corresponds to a field nebula which does not belong to the Coma system, but is only projected on to it. The probability of this assumption is however not very large (1/16). Even if we omit the nebula, the variations in the Coma system still remain very large. In this context it is of interest, to remind the reader that the average density in the Coma cluster is the largest so far observed.

Now, since the relationship between the distance and the redshift is known, we can use the relation to infer the distances of individual nebulae, if their redshifts are spectroscopically measured. Also, we can use it as an independent check of the reliability of the above-mentioned methods of distance determination. To do this, in fact we only need the brightness distribution of all individual nebulae of the same redshift. This new distribution curve must agree with that of Fig. 1, if our original distance determination were correct. This is in fact very approximately the case.

As mentioned above, the redshift means a shift of the entire emission spectrum of the nebula. In addition to the diminution of the apparent photographic magnitude in geometric dependence on the distance, there is still another faintening effect caused by the redshift. The problem of the spatial distribution of the nebulae at large distances is thus not just closely linked with the curvature and absorption of space, but also with the redshift which complicates the whole situation very much.

In the end we must mention some results of van Maanen, which appear to be in contrast with Hubble's determination of distances. Over a period of about twenty years, van Maanen has measured apparent movements (in angular units) of nebulae on the celestial sphere. Since the corresponding angular velocities of the nearest nebulae amount to only about 0.01 arcseconds per year, only nebulae with well-defined, star-like "nuclei" are useful for this purpose, as otherwise the definition of coordinates of the nebula is more difficult due to the blurryness of its photographic image. If one combines van Maanen's angular velocities with Hubble's distances, one obtains extremely high velocities. For NGC 4051, which according to Hubble is located at a distance of 4 million light years, and has an apparent radial velocity of 650 km/s, van Maanen measures an angular velocity of 0.015" per year, which results in a real velocity of 94,000 km/s. This constitutes a big problem. A trivial solution which does not appear impossible from the outset, may be that van Maanen's observed motions do not originate in the nebula, but may be attributed to the reference system used for the stellar background. It must be emphasized however that van Maanen has found similar discrepancies for 13 nebulae. Another result was that all these nebulae appear to move away from the pole of the Milky Way, which seems hard to explain with a movement of the reference system.

Equally important is van Maanen's determination of the rotation of extragalactic spiral nebulae. For Messier 33, according to Hubble at a distance of 900,000 light years, van Maanen observed, superimposed on the above mentioned transversal movement, a rotation of the entire nebula, the components of which for individual objects range from 0.012" up to 0.024" per year. With the mentioned distance, rotation speeds of about 33,000 km/sec result. F.G. Pease measured for NGC 4594 from the Doppler effect at both ends of its diameter a rotation of only 800 km/s. [For Messier 33 itself the observations are not yet completed. However, the rotational velocities are only about 50 km/s.]

If one does not ascribe van Maanen's results to observational errors, but takes these as characteristic of the nebula itself, and one is not ready to drop Hubble's distance determinations, one is facing a serious problem.

## 4. SPECULATIONS CONCERNING THE REDSHIFT.

A complete theory of the redshift must lead to results which meet the following requirements:

1. The redshift is analogous to a Doppler effect, i.e. $\Delta\lambda / \lambda$ for a given nebula is a constant.

2. The apparent Doppler velocity is proportional to the distance $r$ and amounts to 558 km per second per million

parsecs.

3. There is no noticeable absorption and scattering of light in space, which may be associated with the redshift.

4. The definition of the optical images of nebulae is as good as is expected from the resolving power of the instruments. The distance of the objects apparently plays the role expected from geometric considerations.

5. The spectral types of nebulae are essentially independent of distance.

6. The great dispersion of the single values of the radial velocities of the nebulae of dense clusters must be explained in the context of the redshift.

7. The speed of light, on its long way from the nebula to us, is practically the same as the speed of light known to us from terrestrial measurements. This was found from aberration measurements on nebulae by Strömberg and van Biesbroeck.

8. A theory of the redshift, which at the same time does not provide an explanation of van Maanen's results, is at least unsatisfactory.

The facts stated above reflect the observational material up to a distance of about 150 million light years. For their explanation there are presently two general suggestions. The first includes all theories of cosmological character, which are based on the theory of relativity. The second one assumes an interaction of light with matter in the Universe.

## A) Cosmological theories.

In recent years a large number of attempts were made to explain the redshift on the basis of the theory of relativity. Some essential thoughts in this respect are the following.

The general theory of relativity has led to two views regarding the structure of space. The first one is represented by Einstein's quasi-spherical world, while de Sitter has derived the possibility of a hyperbolic space for the case of vanishingly small mass density.

While the geometry of Einstein's space does not lead directly to a redshift, it is necessarily linked with de Sitter's world. However, R. C. Tolman has shown that for the latter case $\Delta\lambda / \lambda$ not only depends on the distance of the nebula, but also on its proper speed. It follows then that apart from the redshift one also has to expect blueshifts which would on average be smaller, but nevertheless of the same order of magnitude as the redshifts, which contradicts the observations. Therefore it was not possible to relate the redshift directly to the curvature of space.

A further important suggestion comes from Friedmann, Tolman, Lemaitre and Eddington, whose work suggests that a static space, according to the theory of relativity, is dynamically unstable, and therefore starts to contract or expand. This result was then interpreted by him [the author does not specify by WHOM] that the redshift would correspond to an actual expansion of the Universe. This proposal has since been discussed by many researchers. The easiest formulation was recently put forward by Einstein and de Sitter (A. Einstein and W. de Sitter, Proc. of the Nat. Acad. Sci., Vol. 18, p. 213, 1932). These two researchers have temporarily given up the existence of an overall curvature of space. The curvature of space was essentially a consequence of the introduction of a so-called cosmological constant $\Lambda$ in Einstein's field equations, which is equivalent to postulate a repulsive force which compensates Newton's attraction for very large distances. This postulate was historically necessary to understand the existence of a non-vanishing mean density which would otherwise lead to infinite gravitational potentials in the limiting case of an infinite static space.

This latter difficulty however disappears automatically, if all matter in space moves away from, or approaches each other. Omitting the cosmological constant $\Lambda$ and the mean curvature, the expansion of matter can then be related directly to the average density. An expansion of 500 km/s per million parsecs, according to Einstein and de Sitter, corresponds to a mean density $\rho \quad 10^{-28}$ g/cm$^3$. Based on observations of self-luminous matter, Hubble estimates $\rho$ $10^{-31}$ g/cm$^3$. It is of course possible that luminous plus dark (cold) matter, taken together, result in a significantly higher density, and the value of ρ    10    g/cm does not therefore appear unreasonable. Einstein's theory further yields the following more precise relationship for the redshift

$$\Delta\lambda/\lambda = \kappa r[1 + 7\Delta\lambda/(4\lambda)]. \tag{3}$$

This means that for large distances the redshift should increase stronger than linearly with the distance. On the basis of the previous observational material it is unfortunately not possible as yet to prove this important conclusion. The most recent observed values of $\Delta\lambda / \lambda$    1/7 for the largest distances are however large enough to expect considerable deviations (25%) from the linear relationship.

Theory also leads to certain conclusions regarding the distribution of brightness levels, number of nebulae, diameter, etc., as function of distance, which however have not yet been proven.

Up to now none of the cosmological theories has dealt with the problem of the large velocity dispersion in dense clusters, such as the Coma system.

**B) Direct influence of existing matter in space on the frequency of light.**

Several years ago I already attempted to consider various physical effects such as the Compton effect on stationary or moving electrons in outer space, the Raman effect, etc., to explain the redshift (F. Zwicky, Proc. Nat. Acad. Sci., Vol. 15, p. 773, 1929). It turned out that none of these can play an important role. When considering effects, which have their origin in an immediate spatial interaction between light and matter, it proves impossible to explain the transparency of intergalactic space.

However, I had then suggested another possible effect, which however will be barely observable on Earth, but for the existence of which some theoretical reasons can be put forward. According to relativity theory, each photon, or light quantum, of frequency {ν} can be assigned an inertial as well as gravitational of $h\nu/c^2$. Thus, there is an interaction (attraction) between light and matter. If the photon is emitted and absorbed at two different points, $P_1$ and $P_2$, respectively, with identical gravitational potentials, then, on the way from $P_1$ to $P_2$, the photon will lose a certain amount of linear momentum and will release it to matter. That photon becomes redder. This effect could be described as gravitational friction, and is caused essentially by the finite velocity of propagation of gravitational effects. Its strength depends on the mean density of matter, as well as on its distribution. In this case the redshift $\Delta\lambda/\lambda$ not only depends on distance, but also on the the distribution of matter. Studies to prove these conclusions are in progress.

In conclusion it has to be said that none of the currently proposed theories is satisfactory. All have been developed on extremely hypothetical foundations, and none of these has allowed to uncover any new physical relationships.

## 5. COMMENTS ON THE VELOCITY DISPERSION IN THE COMA CLUSTER OF NEBULAE

As we have seen in sect. 3, there exist in the Coma cluster apparent differences in velocity of at least 1500 to 2000 km/s. In relation with this enormous velocity dispersion one can make the following considerations.

1. If one assumes that the Coma system has reached a mechanically stationary state, it follows from the Virial theorem

$$\bar{\varepsilon}_k = -\frac{1}{2}\,\bar{\varepsilon}_p \tag{4}$$

where $\bar{\varepsilon}_k$ and $\bar{\varepsilon}_p$ denote the mean kinetic and potential energies, e.g. per unit mass in the system. For the purpose of estimation, we assume that matter is distributed uniformly in the cluster. The cluster has a radius $R$ of approximately one million light years (equal to $10^{24}$ cm) and contains 800 individual nebulae each of a mass of $10^9$ solar masses. The total mass $M$ of the system is therefore

$$M \sim 800 \times 10^9 \times 2 \times 10^{33} = 1.6 \times 10^{45} \text{ g.} \tag{5}$$

From this we have for the total potential energy $\Omega$:

$$\Omega = -\frac{3}{5} \, \Gamma \, \frac{M^2}{R} \tag{6}$$

where $\Gamma$ = gravitational constant

or

$$\bar{\varepsilon}_\mathrm{p} = \Omega/M \sim -64 \times 10^{12} \text{ cm}^2/\text{s}^2 \tag{7}$$

and furthermore

$$\varepsilon_\mathrm{k} = \overline{v^2}/2 = -\bar{\varepsilon}_\mathrm{p}/2 = 32 \times 10^{12} \text{ cm}^2/\text{s}^2 \tag{8}$$
$$(\overline{v^2})^{1/2} = 80 \text{ km/s.}$$

In order to obtain, as observed, a medium-sized Doppler effect of 1000 km/s or more, the average density in the Coma system would have to be at least 400 times greater than that derived on the basis of observations of luminous matter [This would be in approximate accordance with the opinion of Einstein and de Sitter as discussed in Sect. 4.]. If this should be verified, it would lead to the surprising result that dark matter exists in much greater density than luminous matter.

2. One may also assume that the Coma system is not in stationary equilibrium, but that the entire available potential energy appears as kinetic energy. We would then have

$$\varepsilon_\mathrm{k} = -\varepsilon_\mathrm{p} \tag{9}$$

One may thus save only a factor of 2 compared to the assumption 1, and the need for an enormous density of dark matter remains.

3. Let the average density in the Coma cluster be determined purely by luminous matter (*M* as mentioned above). Then the large speeds cannot be explained on the basis of considerations of type 1 or 2 above. If the observed speeds are real anyway, the Coma system should fly apart in the course of time. The end result of this expansion would be 800 single nebulae (field nebulae), which would have proper speeds, as shown in 2., of the order of the original ones (1000 to 2000 km/s). In analogy, one would have to expect that single nebulae with such large proper speeds can also be observed in the present evolutionary state of the Universe. This conclusion hardly matches the experimental facts, as the spread of proper speeds of individually occurring nebulae does not exceed 200 km/s.

4. One may also attempt to consider the speeds as apparent ones, interpreting them as caused by Einstein's redshift. Assuming the above mass *M* one would have for the relative change of the wavelength

$$\Delta\lambda/\lambda \sim -\varepsilon_\mathrm{p}/c^2 \sim 3.5 \times 10^{-8} \,, \tag{10}$$

which is equivalent to a speed of only 10 m/s. Thus, in order to arrive to an explanation of the large velocity dispersion, one would have to permit a much greater density of dark matter than under assuptmions 1 or 2.

These considerations indicate that the large velocity dispersion in the Coma system (and other dense clusters of nebulae) holds an unsolved problem.

# 6. COSMIC RAYS AND REDSHIFTS

Comparing the intensity of visible light from our Milky Way ($L_\mathrm{m}$) with the intensity of light ($L_\mathrm{w}$), that comes to us from the rest of the Universe, one obtains

$$L_\mathrm{m}/L_\mathrm{w} \gg 1 \ . \tag{11}$$

Under the assumption that cosmic rays are not of local nature, one obtains for these the ratio of intensities, $S$, analogous to that in (11)

$$S_\mathrm{m}/S_\mathrm{w} < 0.01 \ . \tag{12}$$

This follows from the practical absence of cosmic ray variations with sidereal time. As I discussed elsewhere (F. Zwicky, Phys. Rev., January 1933) the inequality (12) is difficult to understand because cosmic rays that originate at very large distance, would arrive at Earth with very reduced energy as a result of the redshift. If for example the redshift were consistently proportional to distance, then light quanta from a distance greater than 2000 million light years would reach us with zero energy. (This consideration implies by the way, that even in the presence of an infinite number of stars in the Universe, the light intensity would have a finite, well defined value everywhere.) Under reasonable assumptions about the type of reaction which produces cosmic rays, the inequality (11) is very hard to understand, and the main difficulty lies, as indicated, in the existence of the redshift (F. Zwicky, Phys. Rev., January 1933).

Finally I would like to point out, that the coexistence of the two inequalities (11) and (12) will pose great difficulties for certain recent opinions about the origin of cosmic rays. For example, G. Lemaitre has proposed that one may consider cosmic rays as remnants of certain super-radioactive processes that happened a long time ago. However, at the same time a correspondingly huge amount of visible and ultraviolet light must have been emitted. Since interstellar gases (as well as our atmosphere) absorb cosmic rays more than visible light, the coexistence of inequalities (11) and (12) is incomprehensible.

It is also important, in this context, to point out the following interesting fact. A belt of irregularly shaped boundaries which runs along the Milky Way and stretches from about -10° to about +10° Galactic latitude, completely blocks our view of extragalactic space, i.e. no extragalactic nebula can be observed in this belt. It is known that part of this absorption can be ascribed to very large, dense masses of dust. If cosmic rays were of extragalactic origin, one would actually expect them also to be absorbed along the Milky Way, i.e. one should observe on Earth a variation of cosmic ray intensity with sidereal time. Since such variation is not found, one is tempted to conclude that cosmic rays can not be of extragalactic origin. However, the density and extent of interstellar matter in the Milky Way must be examined more thoroughly.

The present work has emerged from numerous discussions with staff researchers at Mt. Wilson Observatory working in this area. I am particularly indebted to Dr. W. Baade for plenty of valuable advice.

California Institute of Technology, Pasadena.

        El corrimiento al rojo de las nebulosas extragalácticas

    por Fritz Zwicky, Helvetica Physica Acta, Vol. 6, p. 110-127 (1933)

RESUMEN. Este trabajo presenta las características más esenciales de las
nebulosas extragalácticas, así como los métodos utilizados para investigar
estas. En particular, el llamado corrimiento al rojo de las nebulosas
extragalácticas se discute en detalle. Diversas teorías que se han propuesto
para explicar este importante fenómeno, se discuten brevemente. Por último,
se indicará en qué medida el corrimiento al rojo promete ser de importancia
para el estudio de los rayos cósmicos.

1. Introducción

Se ha sabido durante mucho tiempo que hay ciertos objetos en el espacio,
que aparecen, cuando se observa con telescopios pequeños, como muy borrosas
manchas auto-luminosas.  Estos objetos poseen estructuras de diferentes
tipos. A menudo tienen forma esférica, a menudo elíptica, y muchos de ellos
tienen un aspecto como espiral, razón por la cual en ocasiones se les llama
nebulosas espirales. Gracias a la enorme resolución angular de modernos
telescopios gigantes, era posible determinar que estas nebulosas se encuentran
fuera de los límites de nuestra propia Vía Láctea. Imágenes tomadas con el
telescopio de 100 pulgadas de Monte Wilson revelan que estas nebulosas son
sistemas estelares similares al de nuestra propia Vía Láctea.  En general,
las nebulosas extragalácticas se distribuyen uniformemente sobre el cielo y,
como se ha demostrado, también son distribuidas uniformemente en el espacio. Se
presentan como individuos aislados o se agrupan en cúmulos. Las siguientes líneas
intentan hacer un breve resumen de las características más importantes y una
descripción de los métodos que hacen posible establecer estas características.

2. Distancias y características generales de las nebulosas extragalácticas

Como ya se mencionó, se logra, con la ayuda de modernos telescopios, resolver
un buen número de nebulosas total o parcialmente, en estrellas individuales.
En la gran nebulosa de Andrómeda, por ejemplo, un gran número de estrellas
individuales se han observado. Recientemente también cúmulos estelares globulares
han sido descubiertos en esta nebulosa, similar a los que se encuentran dentro
de nuestra propia Vía Láctea. El hecho fortuito de la observación de estrellas
individuales en las nebulosas abre dos maneras de determinar sus distancias.

A) Determinación de la distancia con la ayuda de la relación periodo-brillo
para Cefeidas.

Cefeidas son estrellas cuyo brillo varía periódicamente con el tiempo.
Los períodos están generalmente en el intervalo de uno hasta 60 días. La
magnitud absoluta es función única del período, una función que se ha determinado
para las estrellas de nuestro propio sistema [Vía Láctea]. Por lo tanto, si se
conoce el período, es posible obtener la magnitud absoluta de estas cefeidas
de esta relación. Si, además de eso, uno determina la magnitud aparente y lo
compara con la magnitud absoluta, inmediatamente se obtiene la distancia de las
estrellas. Se han observado varias cefeidas en la nebulosa de Andrómeda. Basado
en esto, la distancia y diámetro de Andrómeda fueron determinados como 900.000
y 42.000 años luz, respectivamente. Para comparación, es importante recordar
que nuestro sistema tiene un límite superior el cual se estima en
unos 100.000 años luz. Las distancias de otras ocho nebulosas se han encontrado
de la misma manera. En las nebulosas más distantes que algunos millones de
años de luz, las cefeidas individuales no pueden resolverse. Por lo tanto,
para determinar su distancia deben ser ideados otros métodos.

B) Estadísticas de las estrellas más brillantes de una nebulosa

Este método se basa en el supuesto de que en los sistemas estelares
extragalácticos la frecuencia relativa de la magnitud absoluta de las estrellas
es la misma como en nuestro propio sistema. La experiencia con los sistemas
vecinos previamente examinados están de hecho de acuerdo con este supuesto.
La magnitud absoluta de estrellas más brillantes de nuestro sistema y sistemas
vecinos resulta ser -6.1 en promedio, con una dispersión de menos de media
magnitud. Aquí sólo notamos que se obtuvieron determinaciones de distancia
similares con la ayuda de Novas.

C) Determinación de distancia de las nebulosas usando su magnitud aparente total.

Con la ayuda de los dos primeros métodos, las distancias de unas sesenta
nebulosas extragalácticas se han encontrado. De la medida del brillo aparente y
de la distancia conocida de estas nebulosas podemos deducir inmediatamente su
brillo absoluto. De esta manera obtenemos la siguiente curva de distribución
(Fig. 1).

Figura 1: H_P magnitud fotográfica absoluta (eje y: número de nebulosas)

La magnitud visual absoluta promedio de las nebulosas es -14.9 con una
dispersión de cerca de cinco magnitudes y un ancho medio de la curva de
distribución de cerca de dos magnitudes. Esta dispersión es, por desgracia,
demasiado grande para permitir una determinación exacta de la distancia a una
nebulosa individual a partir de su brillo aparente y curva de distribución de
magnitud absoluta. Discutimos más adelante cómo es posible todavía determinar
la distancia a ciertas nebulosas individuales con gran precisión. Sin embargo,
el hecho siguiente nos permite encontrar la distancia de un gran número de
nebulosas muy débiles. Como ya se mencionó, las nebulosas a menudo se agrupan en
cúmulos densos, que contienen de 100 a 1000 individuos. Por supuesto que es muy
probable que tal acumulación aparente de nebulosas también es una acumulación
real en el espacio y que por lo tanto todas estas nebulosas están situadas a
aproximadamente la misma distancia. Es relativamente fácil determinar la curva
de distribución de magnitud aparente de las nebulosas de un cúmulo. Esta curva
de distribución es prácticamente la misma que la curva de distribución de la
magnitud absoluta de las sesenta nebulosas, cuyas distancias se han encontrado en
(A) y (B). Esto demuestra que la acumulación aparente de nebulosas [en el cielo]
corresponde a un enjambre denso real en el espacio exterior. Una comparación
del promedio de brillo aparente de las nebulosas en el cúmulo con la magnitud
absoluta promedio de -14.9 inmediatamente nos da la distancia del cúmulo. Las
distancias de los siguientes grupos de nebulosas se determinaron de esta manera.

Tabla 1. Distancia en millones de años luz de varios cúmulos de galaxias
-------------------------
Coma-Virgo        6.
Pegaso           23.6
Piscis           22.8
Cáncer           29.3
Perseo           36.
Coma             45.
Osa Mayor I      72.
Leo             104.
Geminis         135.
-------------------------

El número de nebulosas por unidad de volumen en uno de estos enjambres densos
es por lo menos cien veces mayor que el correpondiente número promedio de
nebulosas individuales dispersadas en el espacio.

Es de interés incluir aquí algunos breves comentarios sobre otras características
de nebulosas que son accesibles a la investigación con la ayuda del telescopio
de 100 pulgadas.

Con respecto a la estructura del Universo, la primera y principal cuestión es
si la distribución de las nebulosas en el espacio es uniforme o no. En el caso
de uniformidad esperamos que el número de nebulosas en una cáscara esférica
de radio r y constante de espesor dr ha de ser proporcional a r^2, siempre
y cuando se trata de un espacio euclidiano. Esta expectación corresponde en
realidad muy exactamente a la realidad, es decir, para la parte del Universo
dentro de alcance de el telescopio de 100 pulgadas. Esto no significa, por
supuesto, que el espacio no podrá finalmente resultar no-euclidiano, una vez
que somos capaces de penetrar más allá en el espacio.

No debemos dejar de mencionar que las conclusiones anteriores sólo son válidas en caso que la absorción y dispersión de la luz en el espacio pueden ser ignoradas. El hallazgo de una distribución uniforme de las nebulosas a las mayores distancias alcanzables con un método que asume la práctica ausencia de absorción y dispersión, es de hecho por sí mismo casi una prueba de esta suposición. En efecto, una distribución uniforme real de las nebulosas sería sesgada por absorción, de tal manera que el número de nebulosas en cáscaras esféricas de espesor constante aumentaría más débilmente con la distancia que $r^2$, y eventualmente disminuirá. En vista de que en efecto la existencia de gases y masas de polvo en el espacio interestelar de nuestro sistema [Vía Láctea] pueden demostrarse, aún así sería de gran importancia tener una prueba independiente de la transparencia del espacio intergaláctico, y mostrar que no es la curvatura del espacio, combinado con la absorción y dispersión, que podría aparentar una distribución uniforme de las nebulosas. Un estudio estadístico de los diámetros aparentes de nebulosas, por ejemplo, respondería a este propósito.

En teoría, la materia intergaláctica debería tener aproximadamente la presión de vapor de los sistemas de estrellas. Suponiendo que el Universo ha llegado un estado estacionario, es posible calcular esta presión (F. Zwicky, Proc. Nat. Acad. Sci., vol. 14, p. 592, 1928). Resulta ser muy pequeña e impediría prácticamente la detección de la materia intergaláctica.

Otra pregunta interesante se relaciona con los tipos espectrales de las nebulosas. La mayor parte de las nebulosas extragalácticas poseen espectros de absorción similares a la del sol con fuertes líneas salientes H y K del calcio y una intensa banda G de Ti (4308 A), Fe (4308 A) y Ca (4308 A). Por lo tanto, las nebulosas pertenecen al tipo espectral G. El tipo espectral es independiente de la distancia, hasta donde las observaciones han llegado en distancia. Un desplazamiento del espectro dependiente de la distancia se discutirá más adelante. La anchura de las líneas de absorción es generalmente de varios Angstroms y también es independiente de la distancia.

Un pequeño porcentaje de las nebulosas observadas también muestra líneas de emisión (Nebulium), generalmente originarios de la región del núcleo de las nebulosas. Lamentablemente muy poco se sabe todavía sobre las condiciones físicas en dichos sistemas.

En tercer lugar es importante investigar la frecuencia relativa de las ya mencionadas diferentes formas de las nebulosas. La distribución estadística es aproximadamente 74% nebulosas espirales, 23% esféricas, y cerca de 3% muestran un aspecto irregular.

En cuarto lugar me gustaría mencionar la determinación de la distribución de la luminosidad dentro de una nebulosa. Esta investigación se ha realizado recientemente por E. Hubble en el Monte Wilson. Hubble obtiene el siguiente resultado preliminar. La luminosidad se puede expresar como una función universal $L(r,alpha)$, donde $r$ es la distancia desde el centro de la nebulosa y $alfa$ es un parámetro adecuado. Mediante una variación de $alfa$, uno puede expresar la distribución de luminosidad en todas las nebulosas con gran precisión (aproximadamente 1%) con la misma función, de hecho hasta valores de $r$, para el cual la luminosidad ha disminuido a 1/1000 de la del centro. También es de importancia con respecto a la práctica falta de absorción y dispersión en el espacio intergaláctico, que la función de distribución de las alfas de las diferentes nebulosas es independiente de la distancia. Por cierto, cabe mencionar que $L$ coincide con la función que corresponde a la distribución de luminosidad en una esfera de gas Emden isotérmica.

En quinto lugar, es de enorme importancia que las nebulosas a gran distancia muestren espectros corridos al rojo, en donde el desplazamiento se incrementa con la distancia. La discusión del llamado corrimiento al rojo es el tema principal del presente trabajo.

3. El corrimiento al rojo de la nebulosa extragaláctica. Relación entre distancia y corrimiento al rojo

V. M. Slipher del observatorio en Flagstaff, Arizona, fue el primero en observar que algunas nebulosas muestran un desplazamiento de sus espectros, que se corresponden a un efecto Doppler de hasta 1800 km/s. Sin embargo, Slipher no estableció ninguna relación entre el corrimiento al rojo y la distancia. Esta relación primero fue sospechada por G. Stroemberg (1925, ApJ 61, 353-388) en su estudio de la velocidad del sol en relación con los objetos más y más distantes. Él encontró que la velocidad media del sol, con relación al sistema de nebulosas vecinas, es grande, del orden de 500 km/s y que el grupo de las nebulosas utilizado muestra una expansión que parece depender de la distancia de la nebulosa individual.

Dado que en el momento de la investigación de Stroemberg no había ninguna
determinación confiable de las distancias de las nebulosas conocidas, K. Lundmark
intentó relacionar la altas velocidades observadas con la compactéz de las
imágenes fotográficas de la nebulosa. Esto se ha demostrado más tarde como un
intento en la dirección correcta. Sin embargo, el intento seguía sin éxito
positivo, ya que se vio después que el diámetro aparente de las nebulosas a
la misma distancia presentan grandes variaciones.

E. Hubble trabajó en Mt. Wilson en la misma dirección. Primero, él también
intentó que se relacionara el corrimiento al rojo con la concentración
aparente de la nebulosa. Aquí partió de la idea de que el desplazamiento al
rojo correspondería al conocido efecto de Einstein. Sin embargo, resultó que
no era posible descubrir relaciones sensatas de esta manera.

En consecuencia, E. Hubble intentó relacionar el corrimiento al rojo con
la distancia de las diferentes nebulosas. Este intento, como es sabido,
desde entonces ha sido de gran éxito. Las nebulosas que estaban disponibles
inicialmente para una investigación tenían distancias desde uno hasta a 6
millones de años luz. La discusión de todos los datos reveló una relación
lineal entre el desplazamiento al rojo y la distancia, con el resultado de
que el corrimiento al rojo corresponde a una velocidad de recesión aparente
de 500 km/s por 1 millón parsec (1 parsec igual a aproximadamente 3.26 años
luz). La dispersión era sin embargo relativamente grande, como por ejemplo la
la galaxia vecina de Andrómeda tiene un desplazamiento al violeta de aprox.
200 km/s, es decir, aparentemente o realmente se mueve hacia nosotros. A pesar de
esto, fue encontrado más tarde que este corrimiento espectral, calculado aquí
por primera vez, fue uno muy bueno. La mejor prueba del increíble cuidado
del trabajo de Hubble es tal vez, que en base a la relación anterior hasta
ahora podía predecir, dentro de pocos porcientos, el desplazamiento al rojo
en cada caso, y de hecho para distancias hasta a treinta veces mayores que las
[muestras de nebulosas] inicialmente usadas.

La dificultad de fotografiar los espectros de las nebulosas muy distantes se
consiste en la necesidad de tiempos de exposición extremadamente largos. De
hecho era necesario exponer placas de hasta cincuenta horas y más y parecían
casi imposible penetrar más en el espacio con este método. En épocas más
recientes grandes avances han sido realizados mediante un espectrógrafo cuyo
lente de cámara tiene una razón focal de f/0.6. Sin embargo, con
ésto se tuvo que sacrificar mucho en la dispersión, y los espectros obtenidos
son solamente cerca de 2 milímetros de largo. Sin embargo, los tiempos de
exposición podrían reducirse hasta unas horas. Sin embargo, no parece
posible penetrar más lejos en el espacio que unos 200 millones de años luz. La
razón de esto radica en parte en la ubicación del telescopio de 100 pulgadas
en las cercanías de la gran ciudad de Los Ángeles, ya que la iluminación del
cielo nocturno y la fuerte dispersión de la luz hacia el telescopio
lamentablemente limita las observaciones astronómicas en el Mt. Wilson, a un
nivel por debajo del rendimiento real del telescopio. Para el telescopio de
200 pulgadas, actualmente en construcción para el California Institute of
Technology se deberá seleccionar una ubicación más adecuada.

Los corrimientos al rojo de varios grupos de nebulosas, expresados como
velocidades de recesión Doppler aparentes, se presentan en la siguiente
tabla 2.

(1) grupo de nebulosas
(2) número de nebulosas en el grupo
(3) diámetro aparente, en grados
(4) distancia en 10^6 años luz
(5) promedio de la velocidad, en km/s

Tabla 2

| (1) | (2) | (3) | (4) | (5) |
|------------|-------|-----|------|-------|
| Virgo | (500) | 12. | 6. | 890 |
| Pegaso | 100 | 1. | 23.6 | 3810 |
| Piscis | 20 | 0.5 | 22.8 | 4630 |
| Cáncer | 150 | 1.5 | 29.3 | 4820 |
| Perseo | 500 | 2.5 | 36. | 5230 |
| Coma | 800 | 1.7 | 45. | 7500 |
| Osa Mayor I | 300 | 0.7 | 72. | 11800 |
| Leo | 400 | 0.6 | 104. | 19600 |
| Geminis | (300) | – | 135. | 23500 |

Estos resultados se graficaron en la Fig. 2.

De esta compilación se desprende que las nebulosas extragalácticas tienen
velocidades que son proporcionales a su distancia. La velocidad específica,
por millones de parsec, es

$$v_s = 558 \text{ km/s} \qquad (1)$$

(véase E. Hubble y M.L.. Humason, ApJ 74, 43, 1931. En este trabajo se puede
encontrar también la bibliografía más esencial.)

El corrimiento al rojo de cada nebulosa individual en promedio se basa en
los corrimientos de al menos tres de las líneas espectrales. Éstos
son generalmente las líneas H y K, la banda G (4303 A) y de vez en cuando
una de las líneas $H_\delta$ (4101 A), $H_\gamma$ (4340 A), Fe (4384 A) y $H_\beta$
(4861 A). La incertidumbre en el corrimiento al rojo del grupo de nebulosas
en Leo resultan de esta manera por ejemplo, en $v = 19621 +-300$ km/s.

Las distintas líneas de absorción sufren el mismo desplazamiento relativo, así
como se espera para el efecto Doppler. Así que tenemos para una cierta nebulosa:

$$\Delta\{lambda\}/\{lambda\} = \text{constante} = K = v/c = \{kappa\}\ r \qquad (2)$$

independientemente de la longitud de onda $\{lambda\}$ y convenientemente el
corrimiento puede expresarse, como lo hemos hecho, en unidades de velocidad. El
mismo valor de K por lo tanto también se aplica al corrimiento máximo del
espectro continuo de emisión.

Hay que tener en cuenta que la figura 1 muestra la media de la velocidad Doppler
de los grupos de las nebulosas. Esta velocidad es el promedio de los valores de
varias nebulosas individuales (de 2 a 9) en cada grupo. Es de gran importanza
para la teoría de los efectos discutidos aquí, que las velocidades de los
miembros individuales de un grupo pueden diferir de la media. En el cúmulo de
Coma, por ejemplo, que hasta ahora es el mejor investigado, se han medido los
siguientes valores individuales.

Velocidades aparentes en el cúmulo de Coma
v = 8500 km/s    6000 km/s
    7900         6700
    7600         6600
    7000         5100(?)

Es posible que el último valor de 5100 km/s corresponde a una nebulosa de
campo que no pertenece al sistema de Coma, pero sólo se proyecta sobre él. La
probabilidad de esta hipótesis sin embargo no es muy grande (1/16). Incluso si
nosotros omitimos la nebulosa, las variaciones en el sistema de Coma siguen
siendo muy grandes. En este contexto resulta de interés, para recordar al
lector que la densidad media en el cúmulo de Coma es la más grande hasta
ahora observada.

Una vez que la relación entre la distancia y el corrimiento al rojo es conocida,
podemos utilizar a la relación para deducir las distancias de las nebulosas
individuales, si sus corrimientos al rojo se miden espectroscópicamente. También,
podemos usarlo como un control independiente de la fiabilidad de los métodos
mencionados anteriormente para la determinación de distancia. Para hacer esto,
de hecho sólo tenemos que determinar la distribución de luminosidad de todas
las nebulosas individuales con el mismo corrimiento al rojo. Esta nueva curva
de distribución debe coincidir con la de la figura 1, si nuestra determinación
de distancia original fuera correcta. Esto es, de hecho, muy aproximadamente
el caso.

Como se mencionó anteriormente, el corrimiento al rojo significa un
desplazamiento de todo el espectro de emisión de la nebulosa. Además de
la disminución de la aparente magnitud fotográfica en dependencia de la
distancia geométrica, hay todavía otro efecto de atenuación causado por el
corrimiento al rojo. El problema de la distribución espacial de las nebulosas a
distancias grandes no está tan sólo estrechamente vinculada con la curvatura y
la absorción del espacio, sino también con el corrimiento al rojo que complica
toda la situación mucho.

Al final hay que destacar algunos resultados de van Maanen, que parecen estar
en contradicción con la determinación de distancias de Hubble. Durante un período
de cerca de veinte años, van Maanen ha medido los movimientos aparentes
(en unidades angulares) de nebulosas en la esfera celeste. Puesto que las
velocidades angulares correspondientes de las nebulosas más cercana asciende
a sólo 0,01 segundos de arco por año, sólo las nebulosas bien definidas, como con núcleos estelares, son útiles para este propósito, ya que la definición de coordenadas de la nebulosa es más difícil debido a la borrosidad de su imagen fotográfica. Si uno combina velocidades angulares de van Maanen con distancias de Hubble, se obtiene velocidades muy altas. Para NGC 4051, que según Hubble se encuentra a una distancia de 4 millones de años-luz, y tiene una velocidad radial aparente de 650 km/s, van Maanen mide una velocidad angular de 0,015" por año, que se traduce en una velocidad verdadera de 94.000 km/s. Esto constituye un gran problema. Una solución trivial que en principio no parece ser imposible, puede ser que los movimientos observados por van Maanen no se originan en la nebulosa, pero pueden ser atribuidos al sistema de referencia utilizado para el fondo estelar. Hay que destacar sin embargo que van Maanen ha encontrado discrepancias similares para 13 nebulosas. Otro resultado fue que estas nebulosas parecen alejarse del polo de la Vía Láctea, que parece difícil de explicar con un movimiento del sistema de referencia.

Igualmente importante es la determinación de van Maanen de la rotación de nebulosas espirales extragalácticas. Messier 33, según Hubble a una distancia de 900000 años-luz, van Maanen observa, sobrepuesto en el movimiento transversal mencionado, una rotación de la nebulosa entera, cuyos componentes para objetos individuales son del orden de 0.012" hasta 0.024" por año. Con la distancia mencionada, resulta una velocidad de rotación de aproximadamente 33000 km/s, mientras que por ejemplo, para NGC 4594 F.G. Pease ha medido, en base al efecto Doppler en ambos extremos de su diámetro, una rotación de sólo 800 km/s. [Para Messier 33 las observaciones no se ha completado. Sin embargo, las velocidades de rotación son sólo cerca de 50 km/s.]

Si uno no atribuye los resultados de van Maanen a errores observacionales, pero toma a estos como característica de la nebulosa en sí misma y uno no está preparado para abandonar las determinaciones de la distancia de Hubble, uno se enfrenta a un grave problema.

4. Especulaciones sobre el corrimiento al rojo.

Una teoría completa del corrimiento al rojo de tiene que llevar a resultados que cumplan los siguientes requisitos.

1. El corrimiento al rojo es análogo a un efecto Doppler, es decir, Delta {lambda} / {lambda} para una nebulosa dada es una constante.

2. La aparente velocidad Doppler es proporcional a la distancia r y asciende a 558 kilómetros por segundo por millón de parsecs.

3. No existe notable absorción y dispersión de la luz en el espacio, que se puede asociar con el corrimiento al rojo.

4. La definición de las imágenes ópticas de las nebulosas es tan buena como lo esperado por el poder de resolución de los instrumentos. La distancia de los objetos aparentemente desempeña el papel esperado por consideraciones geométricas.

5. Los tipos espectrales de las nebulosas son esencialmente independientes de la distancia.

6. La gran dispersión de los valores individuales de las velocidades radiales de las nebulosas de grupos densos deben explicarse en el contexto del corrimiento al rojo.

7. La velocidad de la luz, en su largo camino de la nebulosa a nosotros, es prácticamente la misma velocidad de la luz que conocemos de mediciones terrestres. Esto fue encontrado por mediciones de la aberración en nebulosas por Stroemberg y van Biesbroeck.

8. Una teoría del corrimiento al rojo, que al mismo tiempo no proporciona una explicación de los resultados de van Maanen, es, como mínimo, insatisfactoria.

Los hechos mencionados reflejan el material observado hasta una distancia de unos 150 millones de años-luz. Para su explicación en la actualidad hay dos sugerencias generales. La primera incluye todas las teorías de carácter cosmológico, que se basan en la teoría de la relatividad. En la segunda se asume una interacción de la luz con la materia en el Universo.

A) Teorías cosmológicas.

En los últimos años se hicieron un gran número de intentos para explicar el corrimiento al rojo en base a la teoría de la relatividad. Algunas ideas esenciales en este aspecto son las siguientes.

La teoría general de la relatividad ha dado lugar a dos puntos de vista con respecto a la estructura del espacio. El primero de ellos está representado por el mundo cuasi-esférico de Einstein, mientras que de Sitter se deriva la posibilidad de un espacio hiperbólico para el caso de densidad de masa extremadamente pequeña.

Mientras que la geometría del espacio de Einstein no conduce directamente a un corrimiento al rojo, necesariamente está vinculada con el universo de de Sitter. Sin embargo, R. C. Tolman ha demostrado que para este último caso el valor de Delta {lambda} / {lambda} no sólo depende de la distancia de la nebulosa, sino también de su velocidad propia. Se deduce entonces que, aparte del corrimiento al rojo uno también tiene que esperar un corrimiento al azul que en promedio sería más pequeño, pero sin embargo de la misma magnitud como los corrimientos al rojo, lo cual contradice las observaciones. Por lo tanto no era posible relacionar el corrimiento al rojo directamente a la curvatura del espacio.

Una sugerencia más importante proviene de Friedmann, Tolman, Lemaitre y Eddington, cuyo trabajo sugiere que un espacio estático, según la teoría de la relatividad, es dinámicamente inestable y por lo tanto comienza a contraerse o expandirse. Este resultado fue interpretado entonces por él [el autor no deja claro por QUIEN] que el corrimiento al rojo correspondería a una expansión real del Universo. Esta propuesta ha sido discutido por muchos investigadores. La formulación más sencilla fue dada recientemente por Einstein y de Sitter (Einstein A. y W. de Sitter, Proc. Nat. Acad. Sci., Vol. 18, p. 213, 1932). Estos dos investigadores temporalmente han abandonado la idea de una curvatura global del espacio. La curvatura del espacio fue esencialmente una consecuencia de la introducción de la llamada constante cosmológica {Lambda} en las ecuaciones de campo de Einstein, que es equivalente a postular una fuerza repulsiva que compensa la atracción de Newton para distancias muy grandes. Este postulado era históricamente necesario para comprender la existencia de una densidad media no despreciable que de lo contrario conduce a potenciales gravitatorios infinitos en el caso límite de un espacio estático infinito. Esta última dificultad sin embargo desaparece automáticamente, si todas las masas en el espacio se alejan, o se acercan uno entre los otros. Omitiendo la constante cosmológica {Lambda} y la curvatura media, la expansión de la materia entonces puede ser relacionada directamente con la densidad media. Una expansión de 500 km/s por millón de parsecs, según Einstein y de Sitter, corresponde a una densidad media rho~ 10^-28 g/cm^3. En base a observaciones de la materia auto-luminosa, Hubble estima rho ~ 10^-31 g/cm^3. Por supuesto es posible que la materia luminosa, junto con la oscura (fría), tenga una densidad significativamente mayor, y el valor de rho ~ 10^-28 g/cm^3 por lo tanto, no parece irrazonable.  Además la teoría de Einstein da la siguiente relación más precisa para el corrimiento al rojo

 Delta{lambda}/{lambda} = {kappa} * r [1 + 7 Delta{lambda}/(4{lambda})]. (3)

Esto significa que, para distancias grandes, el corrimiento al rojo debe aumentar más fuerte que linealmente con la distancia. En base al material observado anterior, lamentablemente no es posible todavía demostrar esta importante conclusión. Los valores observados más recientes, de Delta {lambda}/{lambda} ~1/7 para las mayores distancias, son sin embargo suficientemente grandes como para esperar desviaciones considerables (25%) de la relación lineal.

La teoría también conduce a ciertas conclusiones en cuanto a la distribución de brillos, cantidad de nebulosas, diámetro, etc., en función de la distancia, los cuales igualmente aún no se han comprobado.

Hasta ahora ninguna de las teorías cosmológicas ha abordado el problema de la gran dispersión de velocidad en cúmulos densos, tales como el sistema de Coma.

B) Influencia directa de la materia existente en el espacio en la frecuencia de la luz

Hace varios años ya intenté tomar en cuenta varios efectos físicos tales como el efecto Compton en electrones estacionarios o en movimiento en el espacio exterior, el efecto Raman, etc., para explicar el corrimiento al rojo (F. Zwicky, Proc. Nat. Acad. Sci., Vol. 15, p. 773, 1929). Resultó que ninguno de ellos puede jugar un papel importante. Al considerar los efectos, que tienen su origen en una interacción inmediata espacial entre la luz y la materia, resulta imposible explicar la transparencia del espacio intergaláctico.

En cambio, yo había sugerido otro posible efecto, que sin embargo será difícilmente observable en la Tierra, pero para su existencia se pueden presentar

algunas razones teóricas. Según la teoría de la relatividad, cada
fotón, o cuanto de luz, de la frecuencia {nu} puede asignarse una masa tanto
inerte como gravitatoria, de h {nu} / c ^ 2. Por lo tanto, existe una interacción
(atracción) entre la luz y la materia. Si el fotón es emitido y absorbido en
dos diferentes puntos, P1 y P2, respectivamente, con idénticos potenciales
gravitacionales, entonces, en el camino de P1 a P2, el fotón perderá una cierta
cantidad de movimiento [impulso] lineal y lo liberará a la materia. Este fotón
enrojecerá. Este efecto podría ser descrito como fricción gravitacional y
es causada básicamente por la velocidad finita de la propagación de efectos
gravitatorios. La fuerza del efecto depende en la densidad media de la materia, así
como de su distribución. En este caso el corrimiento al rojo Delta {lambda} /
{lambda} depende no sólo de distancia, sino también en la la distribución de
la materia. Estudios para probar estas conclusiones están en curso.

En conclusión hay que decir que ninguna de las teorías actualmente propuestas es
satisfactoria. Todos se han desarrollado sobre bases extremadamente hipotéticas,
y ninguno de éstos ha permitido descubrir nuevas relaciones físicas.

5. Comentarios sobre la dispersión de la velocidad en el cúmulo de Coma de nebulosas

Como hemos visto en la secc. 3, existen en el cúmulo de Coma diferencias evidentes
en la velocidad de al menos 1500 a 2000 km/s. En relación con esta enorme
dispersión de la velocidad se pueden hacer las siguientes consideraciones.

1. Si uno asume que el sistema de Coma ha alcanzado un estado mecánicamente
estacionario, se desprende del teorema Virial que

$$<E\_kin> = -1/2 <E\_pot>$$

donde $<E\_kin>$ y $<E\_pot>$ denotan las energías cinéticas y potenciales medias de,
por ejemplo, una unidad de masa en el sistema. A efectos de valoración, asumimos
que la materia se distribuye uniformemente en el cúmulo. El cúmulo tiene un radio
R de aproximadamente 1 millón de años luz (igual a 10^24 cm) y contiene
800 nebulosas individuales de una masa de 10^9 masas solares. La masa total M
del sistema es por lo tanto

$$M \sim 800 \times 10^9 \times 2 \times 10^{33} = 1.6 \ 10^{45} \ g. \qquad (5)$$

De esto tenemos para la energía potencial total Omega:

$$Omega = -(3/5) \ Gamma \ M^2/R \qquad (6)$$
$$donde \ Gamma = constante \ gravitacional$$

$$o \qquad <E\_pot> = Omega/M \sim -64 \ 10^{12} \ cm^2/s^2 \qquad (7)$$
y además
$$<E\_kin> = <v^2>/2 = -E\_pot/2 = 32 \ 10^{12} \ cm^2/s^2$$
$$sqrt\{<v^2>\} = 80 \ km/s \qquad (8)$$

Con el fin de obtener, como se observa, un efecto Doppler de tamaño medio de
1000 km/s o más, la densidad promedio en el sistema de Coma tendría que ser por
lo menos 400 veces mayor que la derivada en base a observaciones de la materia
luminosa [Esto sería aproximadamente de acuerdo con la opinión de Einstein y
de Sitter como comentamos en la secc. 2.]. Si esto es verificado, conduciría
al resultado sorprendente que la materia oscura existe en una densidad mucho
mayor que la materia luminosa.

2. También uno puede asumir que el sistema de Coma no está en equilibrio
estacionario, pero que la energía potencial disponible total aparece como
energía cinética. Nosotros tendríamos entonces E_kin = - E_pot.

Así uno puede ahorrar sólo un factor de 2 respecto a la hipótesis 1 y la
necesidad de una enorme densidad de materia oscura persiste.

3. Que la densidad media en el cúmulo de Coma esté determinado puramente
por materia luminosa (el valor M arriba mencionado). Entonces las grandes
velocidades no pueden ser explicadas en base a consideraciones del tipo 1
ó 2 anteriores. Si las velocidades observadas son reales de todos modos, el
sistema de Coma debe desintegrarse en el curso del tiempo. El resultado final
de esta expansión serían 800 nebulosas individuales (nebulosas de campo),
que tendrían velocidades, como se muestra en la 2., del mismo orden de los
originales (1000 a 2000 km/s). En analogía, habría que esperar que nebulosas
con tan grandes velocidades propias también se pueden observar en el presente
estado evolutivo del Universo. Esta conclusión no coincide con los hechos
experimentales, que muestra que la dispersión de velocidad de las nebulosas
individuales no supera los 200 km/s.

4. También puede intentarse de considerar las velocidades como aparentes, interpretándolas como debido al corrimiento al rojo de Einstein. Suponiendo la masa anterior M, uno tendría para el correspondiente cambio de la longitud de onda

$$\Delta\{\lambda\}/\{\lambda\} \sim E_{pot}/c^2 \sim 3.5 \times 10^{-8} \qquad (10)$$

que equivale a una velocidad de solamente 10 m/s. Así, con el fin de llegar a una explicación de la dispersión de gran velocidad, uno tendría que permitir una densidad de materia oscura mucho mayor que bajo las suposiciones 1 ó 2.

Estas consideraciones indican que la gran dispersión de velocidad en el sistema de Coma (y otros grupos densos de nebulosas) conlleva un problema no resuelto.

6. Los rayos cósmicos y el corrimiento al rojo

Comparando la intensidad de la luz visible de nuestra Vía Láctea ($L_m$) con la intensidad de la luz ($L_w$), que nos llega del resto del Universo, se obtiene

$$L_m/L_w \gg 1 . \qquad (11).$$

Bajo el supuesto de que los rayos cósmicos no son de carácter local, se obtiene para éstos la relación de intensidades, S, análoga a la de (11)

$$S_m/S_w < 0.01. \qquad (12)$$

Esto se desprende de la práctica ausencia de variaciones de rayos cósmicos con tiempo sideral. Como ya lo comentamos en otra parte (F. Zwicky, Phys Rev., Enero de 1933) la desigualdad (12) es difícil de entender porque los rayos cósmicos se originan a distancias muy grandes, llegando a la tierra con energía muy reducida como resultado del corrimiento al rojo. Si por ejemplo el corrimiento al rojo fuera consistentemente proporcional a la distancia, entonces los cuantos de luz de una distancia mayor que 2000 millones de años luz nos llegarían con cero energía. (Esta consideración implica por cierto, que incluso en presencia de un número infinito de estrellas en el Universo, la intensidad de la luz tendría un valor finito, y bien definida, en todas partes). Bajo supuestos razonables sobre el tipo de reacción que produce rayos cósmicos, la desigualdad (11) es muy difícil de entender y la principal dificultad yace, como se indica, en la existencia del corrimiento al rojo (F. Zwicky, Phys Rev., Enero de 1933).

Por último me gustaría señalar que la coexistencia de las dos desigualdades (11) y (12) plantean grandes dificultades para algunas opiniones recientes sobre el origen de los rayos cósmicos. Por ejemplo, G. Lemaître ha propuesto que se pueden considerar rayos cósmicos como remanentes de ciertos procesos super-radioactivos que sucedieron hace mucho tiempo. Sin embargo, al mismo tiempo una cantidad correspondientemente enorme de luz visible y ultravioleta debio haber sido emitida. Dado que los gases interestelares (al igual como nuestra atmósfera) absorben los rayos cósmicos más que la luz visible, la coexistencia de las desigualdades (1)] y (12) es incomprensible.

También es importante, en este contexto, señalar los siguientes hechos interesantes. Un cinturón de límites de forma irregular que se extiende a lo largo de la Vía Láctea y desde aproximadamente -10{grados} a +10{grados} de latitud Galáctica, bloqueando completamente nuestra visión del espacio extragaláctico, es decir, ninguna nebulosa extragaláctica puede ser observada en esta banda. Se sabe que se puede atribuir parte de esta absorción a masas de polvo muy extendidas y densas. Si los rayos cósmicos son de origen extragaláctico, uno realmente esperaría que también se absorbieran a lo largo de la Vía Láctea, es decir, se debe observar en la Tierra una variación de intensidad de rayos cósmicos con el tiempo sideral. Ya que dicha variación no se encuentra, uno se siente tentado a concluir que los rayos cósmicos no pueden ser de origen extragaláctico. Sin embargo, la densidad y medida de la materia interestelar en la Vía Láctea debe ser examinada más a fondo.

California Institute of Technology, Pasadena.